\RequirePackage{lineno}
\documentclass[fleqn,12pt] {article}
\usepackage{epsf}
\begin{document}
\linenumbers
\title{Development of Richtmyer-Meshkov and Rayleigh-Taylor Instability in presence of magnetic field}
\author{Manoranjan Khan,Labakanta Mandal\thanks{e-mail: labakanta@gmail.com},Rahul Banerjee,Sourav Roy and M. R. Gupta \\
Department of Instrumentation Science \& Centre for Plasma Studies
\\Jadavpur University, Kolkata-700032, India\\}
\date{}
\maketitle

\begin{abstract}
Fluid instabilities like Rayleigh-Taylor,Richtmyer-Meshkov and
Kelvin-Helmholtz instability can occur in a wide range of physical
phenomenon from astrophysical context to Inertial Confinement
Fusion(ICF).Using Layzer's potential flow model, we derive the
analytical expressions of growth rate of bubble and spike for
ideal magnetized fluid in R-T and R-M cases. In presence of
transverse magnetic field the R-M and R-T instability are
suppressed or enhanced depending on the direction of magnetic
pressure and hydrodynamic pressure. Again the interface of two
fluid may oscillate if both the fluids are conducting. However the
magnetic field has no effect in linear case.
\end{abstract}
\newpage

\section*{1. Introduction}

At the two fluid interface if heavier fluid is supported by
lighter fluid, Rayleigh-Taylor instability (RTI) can occur. When a
shock(storng/weak)is passed through the interface of two fluid the
interface will be unstable and Richtmyer-Meshkov instability (RMI)
can occur. The nonlinear structure resemble like a bubble(when the
lighter fluid pushes across the unperturbed interface into the
heavier fluid) and a spike (if the fluids are altered) arise due
to this kind of fluid instability. RTI and RMI can occur from
astrophysical situation to Inertial Confinement Fusion (ICF).

In ICF, high density fuel is compressed and accelerated towards
the origin of the target sphere by multi KJ laser shock. During
shock passage, the interface become unstable which inhibits the
compression in the fusion process. Researchers are searching ways
to stabilize these fluid instabilities. Using Layzer's
approximations several authors[1-3] derive the velocity of bubble
and spike in linear and nonlinear domain. Magnetic fields can also
be generated due to the ponderomotive force when the fluids are
ionized [4,5]. The effect of magnetic field on Rayleigh - Taylor
instability has been studied in details previously by
Chandrasekhar [6]. The growth rate has been found to be lowered
both for continuously accelerated (RTI) and impulsively
accelerated (RMI) two fluid interface when $\vec{k}$ has a
component parallel to the magnetic field [7,8].

 Our paper is addressed to the problem of the temporal
 development of the nonlinear interfacial structure caused by RM
 and RT instability in presence of a magnetic field
 parallel to the surface of separation and perpendicular to the acceleration of the two fluids. The
 wave vector is assumed to lie in the same plane and perpendicular
 to the magnetic field.In this type of geometrical situation,
no effect of the magnetic field is found in the linear
 approximation. However, in the nonlinear regime, the effect of
 magnetic field is predominant. It has been seen that the
 nonlinear growth rate may be enhanced or depressed according as
 the magnetic pressure contribution is either positive or
 negative. We have studied analytically and numerically the non
 linear behavior of the fluid interface in presence of magnetic
 field.

\section*{2. Basic equations and geometry of the problem}
We have considered two infinite fluids of different constant
density separated at $y=0$. The heavier fluid of density $\rho_h$
is along +ve y axis where as the lighter fluid of density $\rho_l$
lies along -ve y axis. Gravity also acts along-ve y direction. The
magnetic field acts along $\hat{z}$ direction. So the
Maxwell equation is easily valid i.e. $\vec{\nabla}.\vec{B}=0$.\\
After perturbation the finger like nonlinear surface is assumed to
be parabolic
\begin{eqnarray} \label{eq:1}
y(x,t)=\eta_0(t)+\eta_2(t)x^2
\end{eqnarray}
\begin{eqnarray}\label{eq:}
\mbox{where}\quad\quad \eta_0>0 \quad \mbox{and} \quad \eta_2<0
\quad\mbox{for
 \nonumber bubble}\quad\qquad\qquad \eta_0<0 \quad
\mbox{and} \quad \eta_2>0 \quad\mbox{spike}
\end{eqnarray}
Again we consider constant fluid density and hence equation of
continuity gives $\vec{\nabla}.\vec{v}=0$, which satisfies the
irrotational fluid motion. Since we are interested the motion the
tip of the bubble, we can neglect the higher order term of
$x^i(i\geq 3)$[9].

Now let us consider the potential function for heavier and lighter
fluid, respectively,

\begin{eqnarray}\label{eq:2}
\phi_h(x,y,t)=a_1(t)\cos{(kx)}e^{-k(y-\eta_0(t))}; \quad y>0
\end{eqnarray}

\begin{eqnarray}\label{eq:3}
\phi_l(x,y,t)=b_0(t)y+b_1(t)\cos{(kx)}e^{k(y-\eta_0(t))}; \quad
y<0
\end{eqnarray}

The fluid motion is governed by the ideal MHD equations

\begin{eqnarray}\label{eq:4}
\rho\left[\frac{\partial \vec{v}}{\partial
t}+(\vec{v}.\vec{\nabla})\vec{v}\right]=-\vec{\nabla}p
-\rho\vec{g}+(\vec{J}\times\vec {B})
\end{eqnarray}

where $v=-\nabla\phi$ and $J=\frac{\nabla\times B}{\mu}$

\begin{eqnarray}\label{eq:5} \mbox{and magnetic induction equation}\quad\frac{\partial
\vec{B}}{\partial t}=\vec{\nabla}\times[\vec{v}\times\vec{B}]
\end{eqnarray}

According to our magnetic field consideration

\begin{eqnarray}\label{eq:6}
\frac{1}{\mu}
(\vec{\nabla}\times\vec{B})\times\vec{B}=\frac{1}{\mu}(\vec{B}.\vec{\nabla})\vec
{B}-\frac{1}{2\mu} \vec{\nabla}(\vec{B^2})
\end{eqnarray}

Using above relations and substituting in $\vec{v}$ in Eq. (4)
followed by use of Eq. (6) leads to Bernoulli's equation for the
MHD fluid

\begin{eqnarray}\label{eq:7}
{\rho}\left[-\frac{\partial \phi}{\partial t}+
\frac{1}{2}(\vec{\nabla}\phi)^2\right]=-{p}-\rho gy-\frac{1}{2\mu}
{B}^2+{f(t)}
\end{eqnarray}

For RM instability the gravitation acceleration g should be
replaced by$g= \Delta u \delta (t)$,where $\Delta u$ is the jump
velocity at the interface and $\delta (t)$ is the delta function.

\section*{3. Kinematical and Dynamical boundary conditions at two fluid interface}

The kinematical boundary conditions are
\begin{eqnarray}\label{eq:9}
\frac{\partial \eta}{\partial t}+(v_h)_{x}\frac{\partial
\eta}{\partial x}=(v_h)_{y}
\end{eqnarray}
\begin{eqnarray}\label{eq:10}
(v_h)_{x}\frac{\partial \eta}{\partial x}-(v_l)_{x}\frac{\partial
\eta}{\partial x}=(v_h)_{y}-(v_l)_{y}
\end{eqnarray}

The Bernoulli's equations for both fluids are
\begin{eqnarray}\label{eq:11} \nonumber\rho_h[-\frac{\partial
\phi_h}{\partial t}+ \frac{1}{2}(\vec{\nabla}
\phi_h)^2]-\rho_l[-\frac{\partial \phi_l}{\partial t}+
\frac{1}{2}(\vec{\nabla}
\phi_l)^2]=-[g(\rho_h-\rho_l)y+(p_h-p_l)\\
+(\frac{{B^2}_{h}}{2\mu_{h}}-\frac{{B^2}_{l}}{2\mu_{l}})]+f_h(t)-f_l(t)
\end{eqnarray}

Further with the help of Eqs. (1) and (2) and the
incompressibility condition $\vec{\nabla}.\vec{v_{h(l)}}=0$, Eq.
(11) simplifies to
\begin{eqnarray}\label{eq:12}
\frac{\partial [\vec{B}_{h(l)}(x,y,t)]}{\partial
t}+(\vec{v}_{h(l)}.\vec{\nabla})\vec{B}_{h(l)}=0
\end{eqnarray}

The above Eqs.[8-11]give the temporal development bubble at the
two fluid interface.

\section*{4. Equation for the structure and instability parameters}

Substituting the $\frac{\partial \eta}{\partial x},\frac{\partial
\eta}{\partial t}$ and expanding the velocity terms in powers of
the transverse coordinate of x and keeping up to $x^2$,we obtain
the following equations

\begin{eqnarray}\label{eq:13}
\frac{d\xi_1}{d\ t}=\xi_3
\end{eqnarray}
\begin{eqnarray}\label{eq:14}
\frac{d\xi_2}{d\ t}=-\frac{1}{2}(6\xi_2+1)\xi_3
\end{eqnarray}
\begin{eqnarray}\label{eq:15}
b_0=-\frac{6\xi_2}{(3\xi_2-\frac{1}{2})}ka_1
\end{eqnarray}
\begin{eqnarray}\label{eq:16}
b_1=\frac{(3\xi_2+\frac{1}{2})}{(3\xi_2-\frac{1}{2})}a_1
\end{eqnarray}
\begin{eqnarray}\label{eq:17}
\xi_1=k\eta_0; \qquad \xi_2=\eta_2/k; \qquad
\xi_3=k^2a_1/\sqrt{kg}
\end{eqnarray}

Where $\xi_1$,$\xi_2$ and $\xi_3$ are nondimensionalized (with
respect to the wave length) displacement,curvature and velocity of
the tip of the bubble.

Now we are interested in magnetic field induction equations.We set
the magnetic field induction equation to satisfy Maxwell's
relation as follows

\begin{eqnarray}\label{eq:18}
B_h(x,y,t)=\beta_{h0}(t)+\beta_{h1}(t)\cos{(kx)}e^{-k(y-\eta_0(t))};
\quad y>0
\end{eqnarray}

for heavier fluid and for lighter fluid

\begin{eqnarray}\label{eq:19}
B_l(x,y,t)=\beta_{l0}(t)+\beta_{l1}(t)\cos{(kx)}e^{k(y-\eta_0(t))};
\end{eqnarray}

Using Eq. (11) and expanding the terms up to $x^2$for both
magnetic field induction equations and we get

\begin{eqnarray}\label{eq:20}
\dot{\beta}_{h0}(t)+(\dot{\beta}_{h1}(t)+\beta_{h1}(t)
k\dot{\eta}_0)\cos{(kx)}e^{-k(y-\eta_0(t))}-k^2a_1\beta_{h1}e^{-2k(y-\eta_0(t))}=0
\end{eqnarray}

for $x^0:$  \quad
${\beta}_{h0}(t)+{\beta}_{h1}(t)=constant=B_{h0}\quad, say$
similarly for lighter fluid

\begin{eqnarray}\label{eq:21}
\beta_{l0}(t)+\beta_{l1}(t)=constant=B_{l0}
\end{eqnarray}

For $x^2:$

\begin{eqnarray}\label{eq:22}
\frac{\delta\dot{B}_{h}}{\delta
B_{h}(t)}=\frac{(\xi_2-\frac{1}{2})}{(\xi_2+\frac{1}{2})}\xi_3;\quad
\delta B_{h}(t)=\frac{\beta_{h}(t)}{B_{h0}}
\end{eqnarray}

which gives

\begin{eqnarray}\label{eq:23}
\delta{B}_{h}(t)=\delta{B}_{h}(t=0)\exp\left[{\int
_0^t\xi_3\frac{(\xi_2-\frac{1}{2})}{(\xi_2+\frac{1}{2})}}d\tau\right]
\end{eqnarray}

Similarly for lighter fluid

\begin{eqnarray}\label{eq:24}
\frac{\delta\dot{B}_{l}}{\delta
B_{l}(t)}=\frac{(\xi_2+\frac{1}{2})}{(\xi_2-\frac{1}{2})}\frac{(\xi_2+\frac{1}{6})}{(\xi_2-\frac{1}{6})}\xi_3;\quad
\delta B_{l}(t)=\frac{\beta_{l}(t)}{B_{l0}}
\end{eqnarray}

Hence

\begin{eqnarray}\label{eq:25} \delta B_{l}(t)=\delta
B_{l}(t=0)exp\left[{\int_0^t\xi_3\frac{(\xi_2+\frac{1}{2})}{(\xi_2-\frac{1}{2})}}\frac{(\xi_2+\frac{1}{6})}{(\xi_2-\frac{1}{6})}d\tau\right]
\end{eqnarray}

so that $ \delta B_{h}(t)<(>0)$ if $ \delta B_{h}(t=0)>(<0)$ and $
\delta B_{l}(t)>0(<0)$ if $ \delta B_{l}(t=0)>(<0)$.

The magnetic pressure difference at the two fluid interface will
be
\begin{eqnarray}\label{eq:26}
\frac{1}{2\mu_{h}}B^2_{h}(x,y,t)-\frac{1}{2\mu_{l}}B^2_{l}(x,y,t)=(\frac{B^2_{h0}}{2\mu_{h}}-\frac{B^2_{l0}}{2\mu_{l}})+k^2\left[\frac{{B^2}_{h0}}{\mu_{h}}\delta
B_{h}(t)(\xi_2+\frac{1}{2}) -\frac{{B^2}_{l0}}{\mu_{l}}\delta
B_{l}(t)(\xi_2-\frac{1}{2})\right]x^2
\end{eqnarray}

Now the Bernoulli's Eq.(10) becomes
\begin{eqnarray}\label{eq:27}
\nonumber\rho_h[-\frac{\partial \phi_h}{\partial t}+
\frac{1}{2}(\vec{\nabla} \phi_h)^2]-\rho_l[-\frac{\partial
\phi_l}{\partial t}+\frac{1}{2}(\vec{\nabla}
\phi_l)^2]=-g(\rho_h-\rho_l)y+k^2\frac{B^2_{h0}}{\mu_h}\delta
B_h(t)(\xi_2+\frac{1}{2})x^2\\
+\frac{B^2_{l0}}{\mu_l}\delta
B_l(t)(\xi_2-\frac{1}{2})x^2+f_h(t)-f_l(t)
\end{eqnarray}
Now we get the following nonlinear equation which describe
temporal development of the tip of the bubble and the velocity of
the bubble

\parbox{11cm}{\begin{eqnarray*}
\frac{d\xi_1}{d \tau}=\xi_3 \hskip 750pt\\
\frac{d\xi_2}{d \tau}=-\frac{1}{2}(6\xi_2+1)\xi_3  \hskip 690pt \\
\frac {\frac{d }{d \tau }{\delta B_h(t)}}{{\delta
B_h(t)}}=\frac{(\xi_2-\frac{1}{2})}{(\xi_2+\frac{1}{2})}\xi_3
\hskip
685pt \\
\frac {\frac{d }{d \tau }{\delta B_l(t)}}{{\delta
B_l(t)}}=\frac{(\xi_2+\frac{1}{2})}{(\xi_2-\frac{1}{2})}
\frac{(\xi_2+\frac{1}{6})}{(\xi_2-\frac{1}{6})}\xi_3 \hskip 645pt \\
\frac{d\xi_3}{d\tau}=-\frac{N(\xi_2,r)}{D(\xi_2,r)}\frac{(\xi_3}{(6\xi_2-1)}+2(r-1)\frac{\xi_2(6\xi_2-1)}{D(\xi_2,r)} \hskip 560pt\\
-\frac{(6\xi_2-1)}{D(\xi_2,r)}[r\frac{k V^2_h}{g}\delta
B_h(t)(2\xi_2+1)+\frac{k V^2_l}{g}\delta B_l(t)(2 \xi _2-1)]
\hskip 500pt
\end{eqnarray*}}\hfill
\parbox{0.1cm}{\begin{eqnarray}\label{eq:36} \end{eqnarray}} \\
\begin{eqnarray}\label{eq:37}
where,\nonumber  \tau=t\sqrt{kg}; \quad r=\frac{\rho
_h}{\rho_l};\quad
D(\xi_2,r)=12(1-r)\xi_{2}^{2}+4(1-r)\xi_{2}+(r+1);
\end{eqnarray}
\begin{eqnarray}\label{eq:38}
N(\xi_2,r)=36(1-r)\xi_{2}^{2}+12(4+r)\xi_{2}+(7-r);
V_{h(l)}=\sqrt{B^2_{h0(l0)}/\rho_{h(l)}\mu _{h(l)}}
\end{eqnarray}

$V_{h(l)}$is the Alfven velocity in the heavier (lighter) fluid.

\section*{5. Asymptotic growth rate}
To find out the asymptotic value of growth rate of bubble we set
$d\xi_2/d\tau=0$ which gives $\xi_2=-1/6$ and at $\tau\rightarrow
\infty$ integrating the last equation of the set of Eq.(27),

For RTI when lighter fluid is conducting:

\begin{eqnarray}\label{eq:41}
[(\xi_3)_{asymp}]_{bubble}=\sqrt{\frac{2Akg}{3(1+A)}}\sqrt{1-2(\frac{1-A}{A})\frac{kV^2_l}{g}[\delta
B_l(\infty)]_{bubble}}
\end{eqnarray}

for spike

\begin{eqnarray}\label{eq:41}
[(\xi_3)_{asymp}]_{spike}=\sqrt{\frac{2Akg}{3(1-A)}}\sqrt{1-2(\frac{1+A}{A})\frac{kV^2_l}{g}[\delta
B_l(\infty)]_{spike}}
\end{eqnarray}

and for RMI when the lighter fluid is conducting, the asymptotic
growth rate is calculated omitting the second part of the last Eq.
of the set of Eq.(27)and integrating,we get

\begin{eqnarray}\label{eq:41}
[(\xi_3)_{asymp}]_{bubble}=\sqrt{\frac{4V^2_l(1-A)}{3(\Delta
u)^2(1+A)}\delta
B_l(\infty)_{bubble}}cot\left[\left\{\frac{3(1+A)}{(1+A)}\sqrt{\frac{4V^2_l(1-A)}{3(\Delta
u)^2(1+A)}\delta B_l(\infty)_{bubble}}\right\}\tau\right]
\end{eqnarray}

for spike

\begin{eqnarray}\label{eq:41}
[(\xi_3)_{asymp}]_{spike}=\sqrt{\frac{4V^2_l(1+A)}{3(\Delta
u)^2(1-A)}\delta
B_l(\infty)_{spike}}cot\left[\left\{\frac{3(1-A)}{(1-A)}\sqrt{\frac{4V^2_l(1+A)}{3(\Delta
u)^2(1-A)}\delta B_l(\infty)_{spike}}\right\}\tau\right]
\end{eqnarray}

\section*{6. Results and discussions}
We have solved the above set of equations using
Runge-Kutta-Fehlberg method to describe the tip of the bubble and
the velocity of the tip of the bubble for different cases.

\textbf{\underline{Case 1}}

Assuming lighter fluid is conducting $B_{l0}\neq 0$ and the
heavier one nonconducting, the hydrodynamic pressure driven force
is suppressed by magnetic pressure.In case of weak shock the RMI
also suppress.For density ratio $\rho_h/\rho_l=r=1.5,$ it has been
seen that RT instability is suppressed (Fig.1).The RM instability
is also suppressed in such situation (Fig.2).If the density ratio
is increased, the Atwood number increases,consequently the growth
rate increases. The growth rate may decrease if the density ratio
is decreased.

\textbf{\underline{Case 2}}

If the heavier fluid is conducting and lighter one nonconducting,
the magnetic pressure acts along +ve y direction which increases
the bubble growth for both case RT(Fig.3) and RM(Fig.4).

\textbf{\underline{Case 3}}

If both the fluids are conducting the magnetic pressure difference
and hydrodynamical pressure difference act in different direction
and also in opposite phase at the interface.Hence the bubble will
shows oscillatory behavior for both cases. For weak shock the
oscillation frequency will be increased with the Alfven
velocity(Fig.5).

\textbf{Applications:}

Super Nova explosion starts in a white dwarf as a laminar
deflagration and RT instability begins to act.In white dwarf,
magnetic field $\sim 10^8 $G at the surface and RT instability
arising during type Ia supernova explosion is associated with the
strong magnetic field[10]. In the solar corona, magnetic field
exist in a range of few Gauss to kilo Gauss. The lower limit of
magnetic field is $\sim$ 10-20 Gauss, having temperature
$2\times10^6$ k[11]. Our model suggest that RT instability may
show oscillatory stabilization if the magnetic field is greater
than 34 gauss in solar corona.

\section * {ACKNOWLEDGEMENTS}
This work is supported by the Department of Science \& Technology,
Government of India under grant no. SR/S2/HEP-007/2008.

\begin{figure}[p]
\vbox{\hskip 1.5cm \epsfxsize=12cm \epsfbox{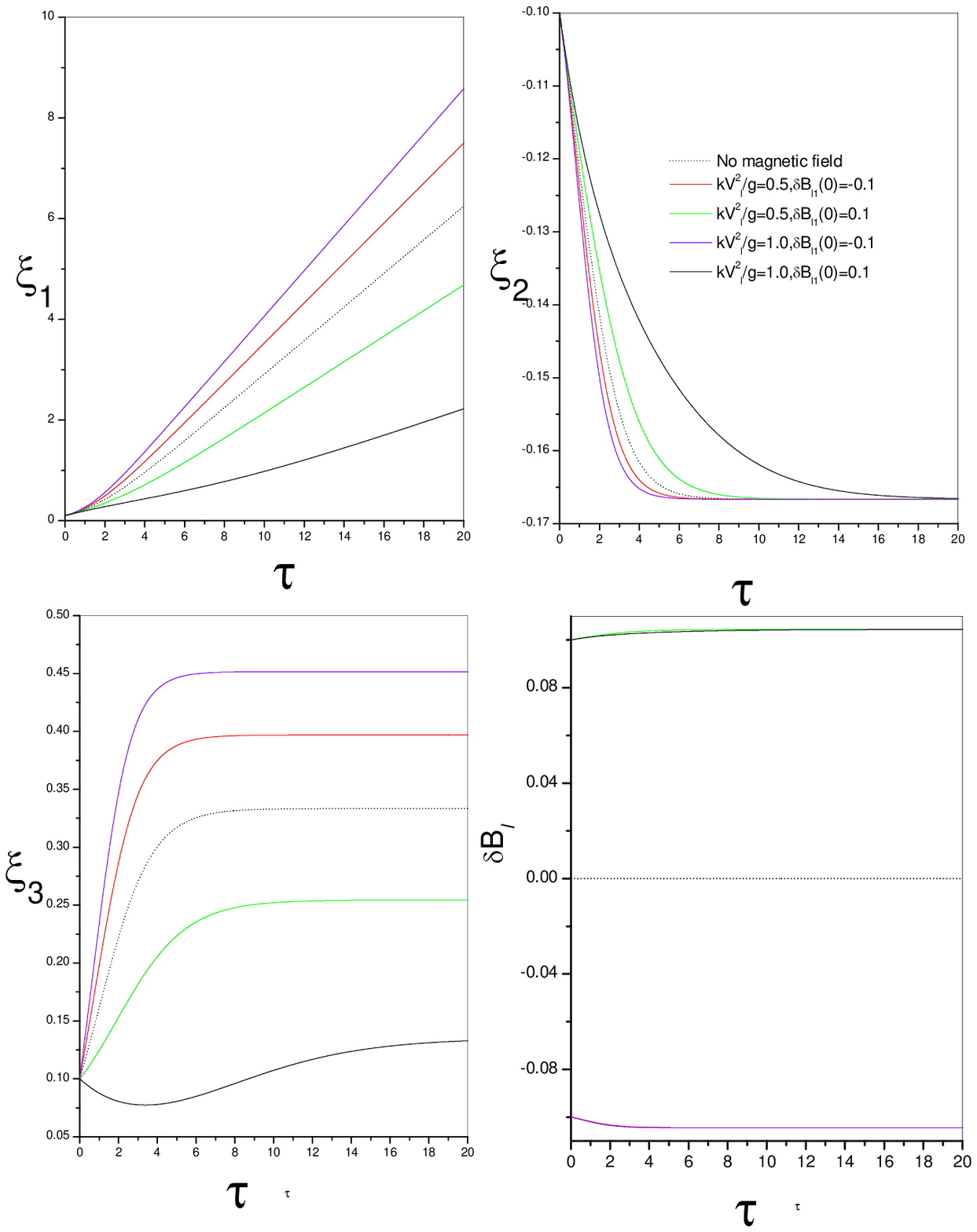}}
\begin{verse}
\vspace{-0.25cm} \caption{Variation of $\xi_1$, $\xi_2$,bubble
growth rate  $\xi_3(=\dot\xi_1)$ and $\delta B_{l}$ with $\tau$
for $V^2_h =0$ for RTI . Initial values are
$\xi_1=0.1,\xi_2=-0.1,\xi_3=0.1$ and $r=1.5$} \label{fig:1}
\end{verse}
\end{figure}

\begin{figure}[p]
\vbox{ \hskip 1.5cm \epsfxsize=12cm \epsfbox{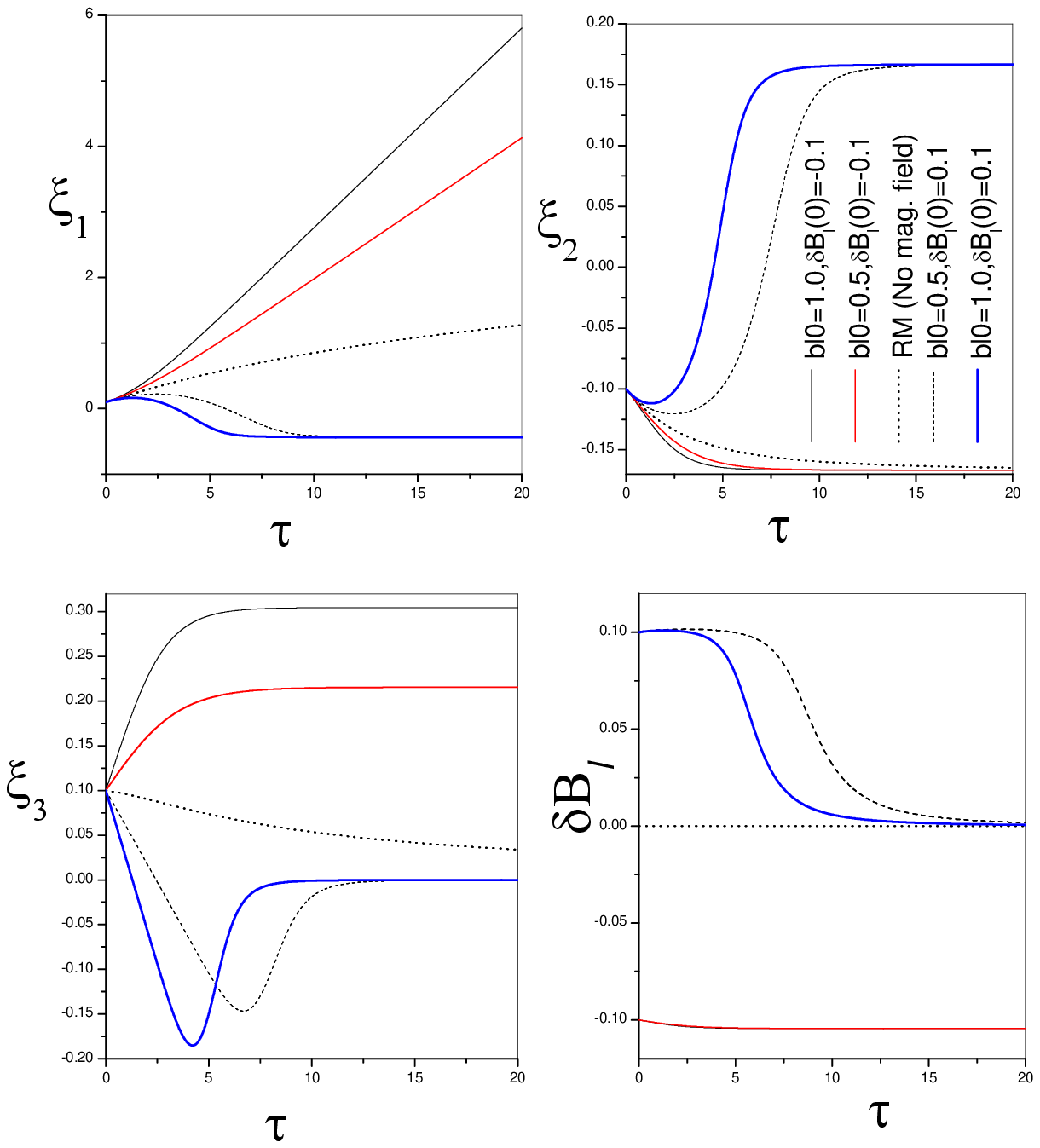}}
\begin{verse}
\vspace{-0.25cm} \caption{Variation of $\xi_1$, $\xi_2$,bubble
growth rate  $\xi_3(=\dot\xi_1)$ and $\delta B_{l}$ with $\tau$
for $V^2_h =0$ for RMI . Initial values are as in fig1.}
\label{fig:2}
\end{verse}
\end{figure}

\begin{figure}[p]
\vbox{ \hskip 1.5cm \epsfxsize=12cm \epsfbox{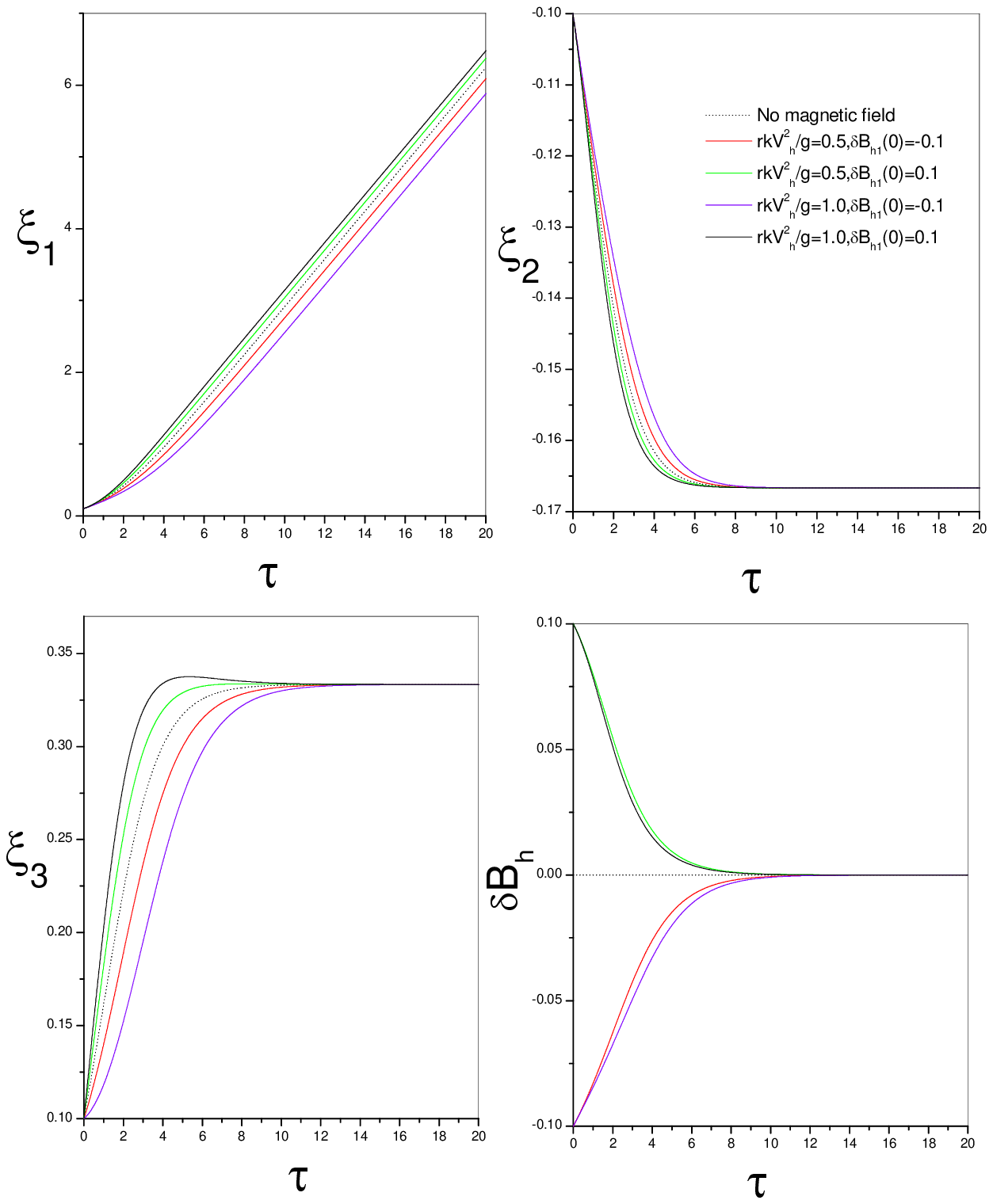}}
\begin{verse}
\vspace{-0.25cm} \caption{Variation of $\xi_1$, $\xi_2$, bubble
growth rate $\xi_3(=\dot\xi_1)$ and $\delta B_{l}$ with $\tau$ for
$V^2_l =0$for RTI .Initial values are as in fig1. } \label{fig:3}
\end{verse}
\end{figure}

\begin{figure}[p]
\vbox{ \hskip 1.5cm \epsfxsize=12cm \epsfbox{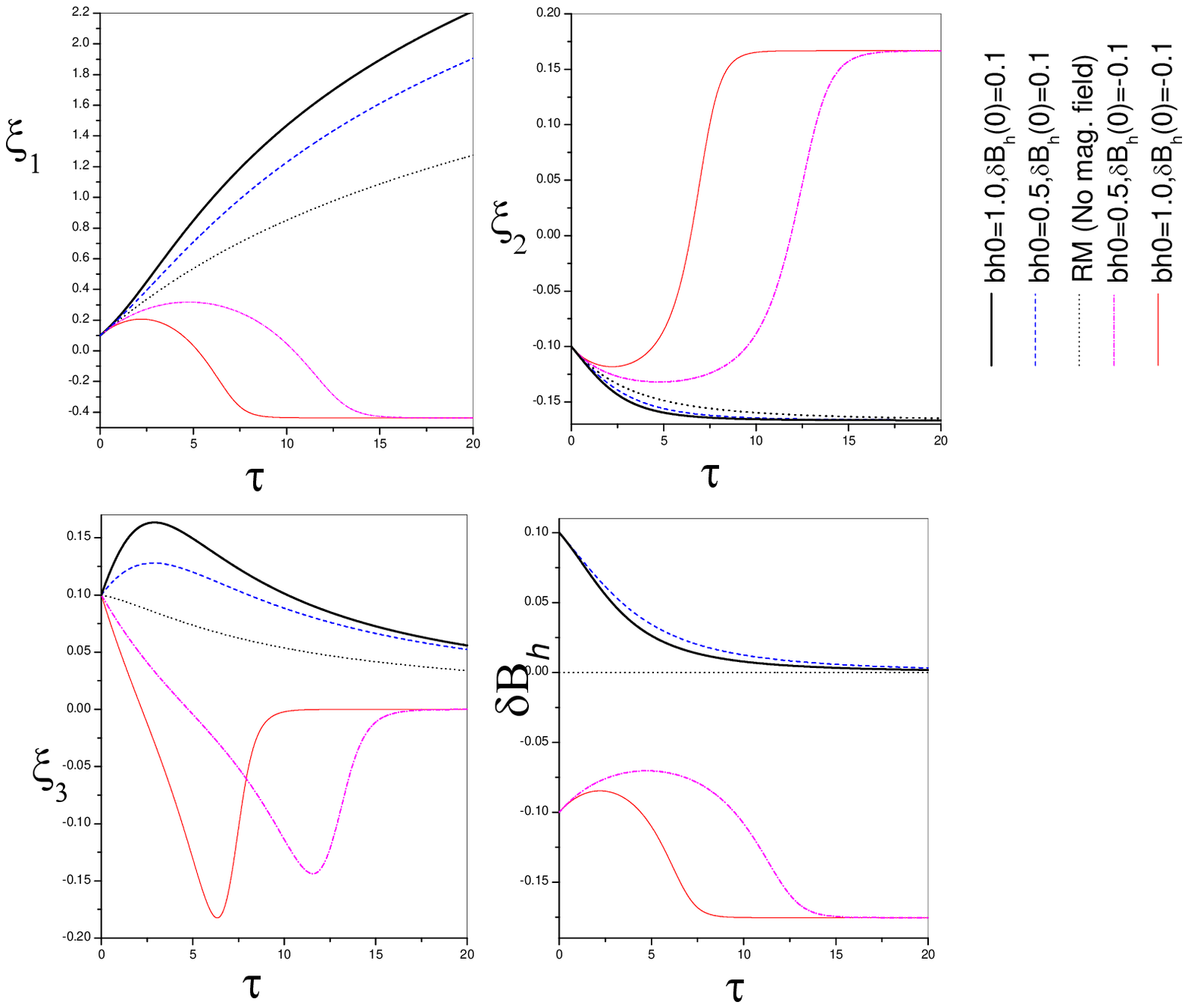}}
\begin{verse}
\vspace{-0.25cm} \caption{Variation of $\xi_1$, $\xi_2$, bubble
growth rate $\xi_3(=\dot\xi_1)$ and $\delta B_{h}$ with $\tau$ for
$V^2_l =0$ for RMI.Initial values are as in fig1. } \label{fig:4}
\end{verse}
\end{figure}

\begin{figure}[p]
\vbox{ \hskip 1.5cm \epsfxsize=12cm \epsfbox{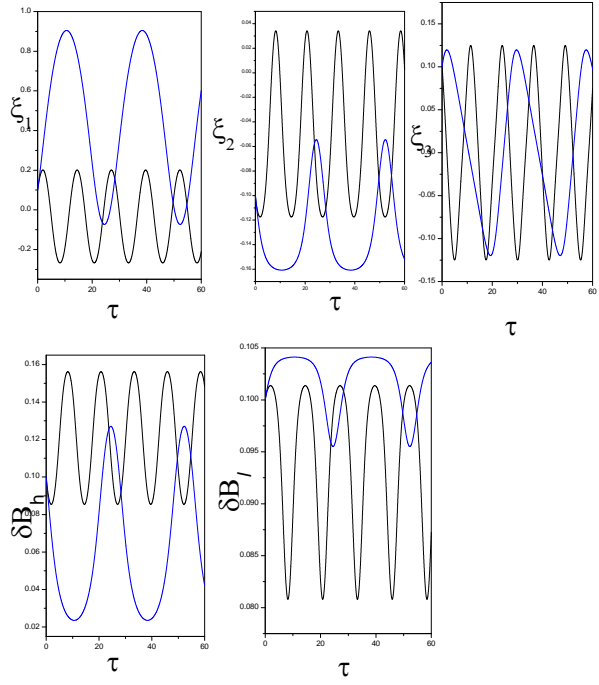}}
\begin{verse}
\vspace{-0.25cm} \caption{Variation of $\xi_1$, $\xi_2$, bubble
growth rate $\xi_3(=\dot\xi_1)$,$\delta B_{h}$ and $\delta
B_{l}$with $\tau$. Initial values
$\xi_1=0.1,\xi_2=-0.1,\xi_3=0.1,\delta B_{h}=\delta
B_{l}=0.1,B_{h0}=B_{l0}=1.5$ and $r=1.5$ blue line for RTI and
black line for RMI.} \label{fig:5}
\end{verse}
\end{figure}

\end{document}